\documentclass[a4paper,12pt]{article}
\pdfoutput=1 

\usepackage{jheppub} 

 \usepackage{amsmath}
\usepackage{graphicx,amssymb,bm,latexsym,amsmath}
\usepackage{amsmath}
\usepackage{epstopdf}
\usepackage{caption}
\usepackage{subcaption}
\usepackage[section]{placeins}
\usepackage{float}
\usepackage{capt-of}
\usepackage{enumitem} 
\usepackage{tabu}

\newcommand{\mathsym}[1]{{}}
\newcommand{\unicode}[1]{{}}

\title{\boldmath Finite Size Effect from Classical Strings in deformed AdS$_3\times$ S$^3$}

\author{}
\author{Kamal L. Panigrahi,}
\author{Manoranjan Samal}
\affiliation{Department of Physics,\\Indian Institute of Technology Kharagpur,\\
	Kharagpur-721 302, India}

\emailAdd{panigrahi@phy.iitkgp.ac.in}
\emailAdd{manoranjan@phy.iitkgp.ac.in}

\abstract{ We study the finite size effect of rigidly rotating and spinning folded strings in $(AdS_3\times S^3)_{\varkappa}$ background. We calculate the leading order exponential corrections to the infinite size dispersion relation of the giant magnon, and single spike solutions. For the spinning folded strings we write the finite size effect in terms of the known Lambert $W$-function.}

\begin{document}
	\maketitle
	\flushbottom
	
	\section{Introduction}
Gauge-gravity duality, particularly the AdS/CFT correspondence \cite{Maldacena:1997re,Witten:1998qj,Gubser:1998bc} has become a major area of research for last few decades in an attempt to understand string theory in terms of gauge theory parameters and vice versa. Appearance of integrability of the planar gauge theory, on one hand, is an interesting theoretical aspect of ${\cal N} = 4$ Supersymmetric Yang-Mills (SYM) and, on other hand, it allows for precision tests of the AdS/CFT correspondence. This very fact has been quite useful for comparing the predictions coming from the string world-sheet sigma model with their counter part spin-chain model in $\mathcal{N}$ = 4 SYM theory. In particular, appearance of integrability on both sides has been quite instructive in establishing the equivalence between the spectrum of conformal dimensions of gauge invariant operators in the gauge theory and the spectrum of the string states in a better way. It was  first observed in \cite{Minahan:2002ve} that the Hamiltonian of Heisenberg spin-chain system is identical to that of certain  gauge invariant operators composed of scalar fields in $\mathcal{N}$=4 SYM theory at one loop level \cite{Beisert:2003yb}.  The conformal dimension of this operator in $\mathcal{N}=4$ SYM theory with large R-charge can be computed using Bethe ansatz equations. Later it was successfully extended to all higher order loops in SYM theory \cite{Beisert:2004hm,Beisert:2005fw}. On the other hand, the type IIB string theory on AdS$_5 \times $ S$^5$ is described by a nonlinear sigma model with a conformal symmetry group PSU($2,2|4$) \cite{Metsaev:1998it}. Though solving the exact sigma model is quite non trivial due to the presence of  infinite number of conserved charges, classical integrability technique provides valuable information on the AdS/CFT duality in the domain of large \rq t Hooft coupling $(\lambda)$ \cite{Bena:2003wd}. \par
	
In this context a large class of semiclassical strings have been studied extensively, for example in \cite{Gubser:2002tv,Minahan:2002rc,Tseytlin:2004xa}. It was first observed by Hofman and Maldecena \cite{Hofman:2006xt} that the scaling relation obeyed by the elementary magnon excitations in spin chains of $\mathcal{N}=4$ SYM theory matches with the scaling relation obtained from certain rigidly rotating strings called  giant magnon  where the linear momentum of the spin chain excitation is identified with certain geometrical angle in the string theory side. It is further generalized to magnon bound state which corresponds to dyonic giant magnon in R$\times$ S$^3$ with two angular momenta \cite{Dorey:2006dq,Chen:2006gea}. It has also been observed that a certain kind of rigidly rotating string called spiky string is dual to higher twist operators in the dual gauge theory side \cite{Kruczenski:2006pk,Kruczenski:2004wg,Ishizeki:2007we}. In addition to the rotating strings, spinning and folded strings, oscillating strings \cite{Tseytlin:2003ii,Arutyunov:2003uj} have also been found out to have exact  dual operators in the gauge theory. \par 
In all the above instances the global charges on both sides of the duality are very large or infinite which makes the structure of the integrability easy to solve. In particular, in case of magnon excitation,  the R-charge of the operator and similarly the angular momentum of its string theory dual are infinites. For such infinite angular momentum cases the asymptotic Bethe ansatz correctly predicts the conformal dimension roughly up to the order of $\lambda^L$ with $L$ the length of the spin chain. Beyond that order, virtual particles start to wrap around the spin chain resulting corrections to the conformal dimension, which is known as wrapping effects \cite{Sieg:2005kd,Ambjorn:2005wa,Kotikov:2007cy}. The modified conformal dimension due to this wrapping effect corresponds to the scaling relation with finite conserved charges in string theory side. So it is quite important to find the spectrum of string states with finite conserved charges and compare with the modified conformal dimension.  In string theory side finite-size  correction to giant magnons or dyonic giant magnons comes in as a term exponentially suppressed in size at classical level as well as quantum level \cite{Arutyunov:2006gs,Astolfi:2007uz,Minahan:2008re,Ramadanovic:2008qd,Klose:2008rx,Shenderovich:2008bs,Ahn:2008gd}. It has been found that the finite size correction for giant magnon  derived from L\"{u}scher formulas of S matrix in strong coupling matches with finite size correction of classical strings. Later it was extended to multi magnon bound states \cite{Hatsuda:2008na}. There are several examples where finite size correction has been computed in string theory side and was compared with their gauge theory counter part, for example  \cite{Ahn:2008sk,Bykov:2008bj,Grignani:2008te,Bozhilov:2010rf,Ahn:2010da,Jain:2008mt}. Some of recent works in this direction can be found in \cite{Floratos:2013cia,Floratos:2014gqa,Ahn:2014tua,Ahn:2014aqa,Ahn:2016egk}. \par
Motivated by above these, in this paper  we wish to find out finite size corrections to classical string solutions in the so called $\varkappa$- deformed background which has been constructed as an integrable background. It is constructed by Yang-Baxter deformation of super currents which has Cartan subgroup $[U(1)]^6$ \cite{Delduc:2013qra,Delduc:2014kha,Hoare:2014pna,Arutyunov:2013ega,Araujo:2017jkb,Araujo:2017jap,Araujo:2017enj}. To explore the integrability beyond the original AdS$_5 \times $ S$^5$ various rigidly rotating string, spinning string and oscillating string have been studied in this deformed background \cite{Kameyama:2014vma,Roychowdhury:2016bsv,Hernandez:2017raj,Barik:2018haz}. The infinite size giant magnon and single spiky string solution in $\varkappa$-deformed background have been studied in \cite{Arutynov:2014ota,Khouchen:2014kaa,Banerjee:2014bca}. It is quite important to compute the exponentially suppressed finite size corrections of these sting solutions.  For the giant magnon in the $\eta$-deformed background \footnote{${\varkappa = \frac{2\eta}{1-\eta^2}}, ~{\rm where}~\varkappa \in [0, \infty)$}, the finite size effect has been calculated in \cite{Ahn:2014aqa,Ahn:2016egk}. \par
The rest of the paper is organized as follows, In section 2, we study rigidly rotating strings which includes both giant magnon and single spike string solutions in $\varkappa$-deformed R$\times$S$^2$ background. Taking the relevant ansatz, we find the equations of motion and the corresponding  string solutions. Then we compute the finite size corrections to the scaling relation by expanding the charges at large values. Section 3 is devoted to the study of finite size correction in the leading oder of the charges for the  spinning folded string and the result is written in terms of Lambert W-function. In section 4, we conclude with some remarks.
	
\section{Finite size rigidly rotating strings}	
In this section, we present the finite size expansion of giant magnon and single spike strings in $\varkappa$-deformed background in the leading order expansion of the large charges. We start with the truncated R $\times$ S$^2$ metric of the full $(\text{AdS}_5 \times \text{S}^5 )_\varkappa$   deformed geometry
\begin{equation}
ds^2=-dt^2+\frac{d\theta^2}{1+\varkappa^2 \cos^2 \theta}+\frac{\sin^2 \theta d \phi^2}{1+\varkappa^2 \cos^2 \theta},
\end{equation}
where $\varkappa \in (0,\infty]$ is the deformation parameter. The Polyakov action for a string in this background in conformal gauge is given  by 
\begin{equation}
S=\frac{\hat{T}}{2} \int d \tau d \sigma \left[-(\dot{t}^2-t'^2)+\frac{\dot{\theta}^2-\theta'^2}{1+\varkappa^2 \cos^2 \theta}+\sin^2 \theta \frac{\dot{\phi}^2-\phi'^2}{1+\varkappa^2 \cos^2 \theta}\right],
\end{equation}
where `dot' and `prime' denote the derivative with respect to $\tau$ and $\sigma$ respectively and $\hat{T}$ is the effective string tension, $\hat{T} = \frac{\sqrt{\lambda}}{2 \pi}\sqrt{1+\varkappa^2}$. The equations of motion that follow from the above action are to be supplemented by the following Virasoro constraints
\begin{eqnarray}
g_{\mu \nu}\partial_\tau X^\mu \partial_\sigma X^\nu=0, \nonumber \\
g_{\mu \nu}\left(\partial_\tau X^\mu \partial_\tau X^\nu+\partial_\sigma X^\mu \partial_\sigma X^\nu\right)=0,
\end{eqnarray}
where $g_{\mu \nu}$ is the space-time metric and $X^\mu=\{t, \theta, \phi\}$. We take the following ansatz for  rigidly rotating string 
	\begin{equation}
	t=\tau,\quad \theta=\theta(y), \quad \phi= \alpha \tau+h(y),
	\end{equation}
	where $y= \sigma-\beta \tau$.
The equation of motion of $\theta$ takes the form 
\begin{equation}\label{eomtheta}
\frac{ \theta_{yy}}{1+\varkappa^2 \cos^2 \theta}+\frac{\varkappa^2\theta_y^2  \sin 2\theta}{2(1+\varkappa^2 \cos^2 \theta)^2}-\frac{\left( (\alpha-\beta h_y)^2-h_y^2 \right)\sin 2 \theta}{2(\beta^2-1)(1+\varkappa^2 \cos^2 \theta)} \left(1+\frac{\varkappa^2 \sin^2 \theta }{1+\varkappa^2 \cos^2 \theta}\right)=0,
\end{equation}
where the subscript $y$ denotes derivative with respect to $y$.
	 Similarly the equation motion of $\phi$ gives 
		\begin{equation}\label{phieqn}
	\frac{\partial h}{\partial y}= \frac{\alpha \beta+ A\left(\frac{\varkappa^2 \cos^2 \theta+1}{\sin^2 \theta}\right)}{\beta^2-1},
	\end{equation}
	where $A$  is an integration constant. From the second Virasoro constraint equation we find 
	\begin{equation}\label{thetaeqn}
	\frac{\partial \theta}{\partial y}=\pm B\frac{\sqrt{(\cos^2 \theta_0-\cos^2 \theta)(\cos^2 \theta- \cos^2 \theta_1)}}{(1-\beta^2) \sin \theta},
	\end{equation}
	where 
	$$\cos^2 \theta_0=\frac{\alpha^2-1}{\alpha^2+\varkappa^2}, \quad   \cos^2 \theta_1=\frac{\alpha^2-\beta^2}{\alpha^2+\varkappa^2 \beta^2} \quad \text{and} \quad B=\frac{\sqrt{(\alpha^2+\varkappa^2)(\alpha^2+\varkappa^2 \beta^2)}}{\alpha}.$$
	Substituting (\ref{phieqn}) and (\ref{thetaeqn}) in the first Virasoro constraint equation gives
	\begin{equation}\label{intconstant}
	A=-\frac{\beta}{\alpha}.
	\end{equation}
	It can be easily checked that the equations of motion for $\theta$ and $\phi$ are consistent with the Virasoro constraint equations.
	Explicitly, if one substitutes eqn. (\ref{phieqn}) and eqn. (\ref{thetaeqn}) along with the integration constant (\ref{intconstant}), the eqn. (\ref{eomtheta}) is satisfied. Now we calculate the conserved charges, namely the total energy $\text{E}$, the angular momentum $\text{J}$ and the angular difference $\Delta \phi$ between the end point of the string as   
	 \begin{eqnarray}
	 \text{E}&=& 2 \hat{T}\int_{\theta_{min}}^{\theta_{max}}\frac{d \theta}{\theta'} \frac{\partial \mathcal{L}}{\partial \dot{t}} \nonumber \\
	 &=&   \frac{2 \hat{T}(1-\beta^2)}{B} \int_{\theta_{min}}^{\theta_{max}} \frac{d \theta \sin \theta}{\sqrt{(\cos^2 \theta_0-\cos^2 \theta)(\cos^2 \theta- \cos^2 \theta_1)}}, \\	 \label{magnon energy exp}
	 	 \text{J} &=&  2 \hat{T}\int_{\theta_{min}}^{\theta_{max}}\frac{d \theta}{\theta'} \frac{\partial \mathcal{L}}{\partial \dot{ \phi}} \nonumber \\ 
	 &=& \frac{2 \hat{T}(1-\beta^2)}{B} \int_{\theta_{min}}^{\theta_{max}} \frac{d \theta \sin^3 \theta }{\sqrt{(\cos^2 \theta_0-\cos^2 \theta)(\cos^2 \theta- \cos^2 \theta_1)}} \frac{(\alpha-\beta h_y)}{1+\varkappa^2 \cos^2 \theta}, \\ \label{magnon angular momentum exp}
	 \Delta \phi &=& 2 \int_{\theta_{min}}^{\theta_{max}} \frac{d \theta}{\theta_y} h_y \nonumber \\
	 &=& \frac{2 (1-\beta^2)}{B} \int_{\theta_{min}}^{\theta_{max}} \frac{d \theta \sin \theta}{\sqrt{(\cos^2 \theta_0-\cos^2 \theta)(\cos^2 \theta- \cos^2 \theta_1)}} h_y.  \label{magnon momentum exp}
	 \end{eqnarray}
	 
	 Now depending on the values of parameters $\alpha$ and $\beta$ we have two interesting class of solutions
	 \begin{enumerate}[label=(\roman*)]
	 	\item For $\alpha \geq 1$ and $\beta\leq 1$  : Giant magnon
	 	 \item For $\alpha \geq 1 $ and $\beta \geq 1 $ with $\frac{\beta}{\alpha} \leq 1$ : Single spike
	 \end{enumerate} 
 Below we are going to derive the finite size corrections to these class of solutions in detail.
	\subsection{Giant Magnon solution}
	 Taking $\cos \theta= z$ and $\varepsilon=\frac{z_{min}^2}{z_{max}^2}$, we can check that for $\alpha \geq 1$ and $\beta\leq 1$ 
	 	\begin{equation}
	 z_{max}=\cos^2\theta_1 \quad \text{and} \quad z_{min}=\cos^2\theta_0.
	 \end{equation}
	  Now integrating  (\ref{phieqn}) and  (\ref{thetaeqn}) we get the following solutions of $\theta$ and $\phi$ respectively 
	 \begin{equation}
	 z=\cos\theta=z_{max} \text{dn}\left[\frac{(\sigma- \beta \tau) B z_{max}}{1-\beta^2},1-\varepsilon\right],
	 \end{equation}
	 \begin{multline}
	 \phi = \alpha \tau+ \frac{\beta \sqrt{\alpha^2+ \varkappa^2}}{\sqrt{\alpha^2-\beta^2}} \mathbf{F} \left[\sin^{-1}\left(\frac{1}{\sqrt{1-\varepsilon}} \sqrt{1-\frac{z^2}{z_{max}^2}}\right),1-\varepsilon\right]  \\ - \frac{\alpha^2+\beta^2 \varkappa^2}{\beta \sqrt{\alpha^2-\beta^2}\sqrt{\alpha^2+\varkappa^2}} \mathbf{\Pi}\left[  \frac{z_{min}^2-z_{max}^2}{1- z_{max}^2},\sin^{-1}\left(\frac{1}{\sqrt{1-\varepsilon}} \sqrt{1-\frac{z^2}{z_{max}^2}}\right),1-\varepsilon\right].	 
	 \end{multline}
	 Figure:\ref{fig:giant magnon and spiky string} depicts the snapshots of above giant magnon solution for different values of $\alpha$ with $\beta=0.8$ on the deformed $S^2$. For $\alpha=1$ the above solutions reduce to giant magnon solution with infinite charges, where the two ends of the string lie on equator as depicted in figure:\ref{fig:giant magnon and spiky string}. 
The total energy, angular momentum and angular deficit between two end points of the magnon can be written in elliptic integrals as
	\begin{eqnarray}
	\mathcal{E}=\frac{\pi\text{E}}{\sqrt{\lambda}}= \frac{\sqrt{1+\varkappa^2}(1-\varepsilon)z_{max}}{\sqrt{(1-\varepsilon z_{max}^2)(1+\varkappa^2 z_{max}^2 )}} \mathbf{K}(1-\varepsilon),
	\end{eqnarray}
	\begin{eqnarray}
\mathcal{J}=\frac{\pi\text{J}}{\sqrt{\lambda}}=  \frac{\sqrt{1+\varkappa^2}\sqrt{1+\varepsilon\varkappa^2 z_{max}^2}}{z_{max}\varkappa^2 \sqrt{1+\varkappa^2 z_{max}^2}}\left(\mathbf{\Pi}\left[\frac{\varkappa^2z_{max}^2(1-\varepsilon)}{1+\varkappa^2 z_{max}^2},1-\varepsilon \right]-\mathbf{K}(1-\varepsilon)\right), \nonumber \\
	\end{eqnarray}
	\begin{eqnarray}
	\Delta \phi = p = \frac{2\left((1-z_{max}^2)\mathbf{K}(1-\varepsilon)-(1-\varepsilon z_{max}^2)\mathbf{\Pi}\left[\frac{(1-\varepsilon) z_{max}^2}{z_{max}^2-1},1-\varepsilon\right]\right)}{z_{max}\sqrt{z_{max}^2-1}\sqrt{\varepsilon z_{max}^2-1}}.
	\end{eqnarray}
	In case of giant magnon solution, $\theta_0$ is exactly $\frac{\pi}{2}$ for which both the energy and angular momentum become infinity whereas the difference between them remains finite as shown in appendix A for completeness. To obtain the finite size corrections to the scaling relation we take $\theta_0 \rightarrow \frac{\pi}{2}$ which yields $\varepsilon \rightarrow 0$. First we express $z_{max}$ in terms of $\varepsilon$, $p$ and $\mathcal{J}$ which can be achieved by expanding $p$  around both $\varepsilon=0$ and $z_{max}= \sin \frac{p}{2}$ and subsequently inverting the series for $z_{max}(\varepsilon,p,\mathcal{J})$. Then we insert $z_{max}$ in the expansion of angular momentum at $\varepsilon=0$ and  use the Lagrange inversion formula to get the analytic function $\mathcal{\varepsilon}(p, \mathcal{J})$. Finally we plug $z_{max}(\varepsilon,p,\mathcal{J})$ and $\mathcal{\varepsilon}(p, \mathcal{J})$  in the expression of $\mathcal{E}- \mathcal{J}$  to get the expansion of the scaling relation for finite size corrections.\par
	Before the expansions of $p$ and $\mathcal{J}$ we can see there are logarithmic singularities in Jacobi elliptic function $\mathbf{\Pi}$ which can be isolated using the addition formula (\ref{addition formula}) for which  $\mathcal{J}$ and $\Delta \phi$ turn out to be
	
	\begin{equation}\label{modified angular momentum}
	\mathcal{J}=\sqrt{1+\varkappa^2}\left(\frac{i \pi \mathbf{F}[\sinh^{-1} \varkappa z_{max},\varepsilon]}{2\varkappa\mathbf{K}[\varepsilon]}
	+\frac{ z_{max}(1-\varepsilon)\mathbf{K}(1-\varepsilon)}{\sqrt{(1+\varkappa^2 z_{max}^2)(1+\varepsilon\varkappa^2 z_{max}^2)}}\frac{\mathbf{\Pi}\left[\frac{\varepsilon(1+\varkappa^2z_{max}^2)}{1+\varepsilon\varkappa^2z_{max}^2},\varepsilon\right]}{\mathbf{K}(\varepsilon)}\right),
	\end{equation}
	
	\begin{equation}\label{modifiedp}
	 p = \frac{\pi \mathbf{F}[\sin^{-1} z_{max},\varepsilon]}{\mathbf{K}[\varepsilon]}-\frac{2 z_{max}(1-\varepsilon)\mathbf{K}(1-\varepsilon)}{\sqrt{(1-z_{max}^2)(1-\varepsilon z_{max}^2)}\mathbf{K}(\varepsilon)}\left(\mathbf{K}(\varepsilon)-\mathbf{\Pi}\left[\frac{\varepsilon(1-z_{max}^2)}{1-\varepsilon z_{max}^2},\varepsilon\right]\right).
	\end{equation}
	Similarly the linear momentum $p$ has logarithmic singularity due to Jacobi elliptic function $\mathbf{K}(1-\varepsilon)$
	\begin{equation}
	\mathbf{K}(1-\varepsilon)=\sum_{n=0}^\infty \varepsilon^n \left(d_n \ln \varepsilon +h_n\right)
	\end{equation}
	where $$ d_n=-\frac{1}{2} \left(\frac{(2n-1)!!}{(2n)!!}\right)^2, \quad h_n= -4 d_n(\ln2 + H_n-H_{2n})  ~~ \text{ with} ~~ H_n=\sum_{k=1}^n \frac{1}{k}.$$
	To remove the logarithmic singularities from $p$, we solve  for $\mathbf{K}(1-\varepsilon)$ in angular momentum  and substitute in (\ref{modifiedp}). Then we expand it around both $\varepsilon=0$ and $z_{max}= \sin \frac{p}{2}$. After the expansion we invert the series for $z_{max}$ up to first order of $\varepsilon$ which gives 
	\begin{equation}
	z_{max}=\sin\frac{p}{2}+\frac{1}{8}\cos^2 \frac{p}{2}\sqrt{1+\varkappa^2 \sin^2 \frac{p}{2}}\left(\frac{4\mathcal{J}}{\sqrt{1+\varkappa^2}}+\frac{4 \sinh^{-1}(\varkappa \sin \frac{p}{2})}{\varkappa}+\frac{2 \sin \frac{p}{2}}{\sqrt{1+\varkappa^2 \sin^2 \frac{p}{2}}}\right)\varepsilon.
	\end{equation}
	Now we substitute above $z_{max}(\varepsilon,p,\mathcal{J})$ series in (\ref{modified angular momentum}) and expand around $\varepsilon=0$ which can be written in the form of 
	\begin{equation}
	\mathcal{J}=\sum_{n=0}^\infty \varepsilon^n(c_n \ln \varepsilon+ b_n) \Rightarrow \ln \varepsilon = \frac{\mathcal{J}- \sum_{n=0}^\infty b_n \varepsilon^n}{\sum_{n=0} c_n \varepsilon^n}.
	\end{equation}
	After exponentiating and rearranging it, we get
	
	\begin{equation}\label{giante0exp}
	\varepsilon_0=  \varepsilon \exp \left(\sum_{n=0}^\infty b_n \varepsilon^n -\left(\frac{\mathcal{J}-b_0}{c_0}-\sum_{n=0}^\infty \frac{b_n}{c_0} \varepsilon^n\right)\sum_{n=1}^\infty (-1)^n \left(\sum_{k=1}^\infty \frac{c_k}{c_0}\varepsilon^k\right)^n\right)
	\end{equation}
	where $$\varepsilon_0= \exp \left[\frac{\mathcal{J}-b_0}{c_0} \right]=16 \exp \left(- \frac{2\left( \mathcal{J}\varkappa+\sqrt{1+\varkappa^2} \sinh^{-1} (\varkappa \sin\frac{p}{2})\right)\sqrt{1+\varkappa^2\sin^2\frac{p}{2}}}{\varkappa\sqrt{1+\varkappa^2}\sin\frac{p}{2}}\right) $$ is the lowest order power in (\ref{giante0exp}). 
	Now inverting the above series for $\varepsilon$ we get the general form of $\varepsilon$ as
	\begin{equation}
	\varepsilon= \sum_{n=1}^ \infty \sum_{m=0}^{2n-2} \mathcal{A}_{nm} ~\mathcal{J}^m ~\varepsilon_0^n
	\end{equation} and the leading order terms can be found as
	
	\begin{equation}
	\varepsilon_{leading}= \sum_{n=1}^ \infty  \mathcal{A}_{n, 2n-2} ~\mathcal{J}^{2n-2} ~\varepsilon_0^n= \varepsilon_0+\frac{256\cot^2\frac{p}{2}}{1+\varkappa^2} \mathcal{J}^2 \varepsilon_0^2+\mathcal{O}(\mathcal{J}^4 \varepsilon_0^3).
	\end{equation}
	Finally substituting this in the expression of $(\mathcal{E}-\mathcal{J})$ gives the leading finite size dispersion relation for giant magnon
	\begin{eqnarray}\label{GM finite size correction}
	\mathcal{E}-\mathcal{J}| _{leading}=\sqrt{1+\varkappa^2}\left[\frac{\sinh^{-1} (\varkappa\sin \frac{p}{2})}{\varkappa}-\frac{(1+\varkappa^2 )\sin^3 \frac{p}{2}}{4\sqrt{1+\varkappa^2 \sin^2 \frac{p}{2}}} \varepsilon_0-\frac{ \csc \frac{p}{2} \sin^2 p }{32\sqrt{1+\varkappa^2 \sin^2 \frac{p}{2}}} \mathcal{J}^2 \varepsilon_0^2\right]. \nonumber \\
	\end{eqnarray}
	We can see, for $J \rightarrow \infty$ in the above result  gives
	\begin{equation}
	\mathcal{E}-\mathcal{J}=\sqrt{1+\varkappa^2}\frac{\sinh^{-1} (\varkappa\sin \frac{p}{2})}{\varkappa}
	\end{equation}
	which matches with the infinite size giant magnon dispersion relation \cite{Arutynov:2014ota,Khouchen:2014kaa}. Similarly for $\varkappa \rightarrow 0$ it reduces to the finite size dispersion relation of giant magnon in original R$_t\times$ S$^2$ \cite{Hofman:2006xt}. This result also exactly matches with finite size correction for giant magnon  obtained from dressing phase at strong coupling \cite{Ahn:2016egk}.

%

	\subsection{Single Spike solution}
	
\begin{figure}[H]
		\centering
		\begin{subfigure}{.5\textwidth}
			\centering
			\includegraphics[width=.7\linewidth]{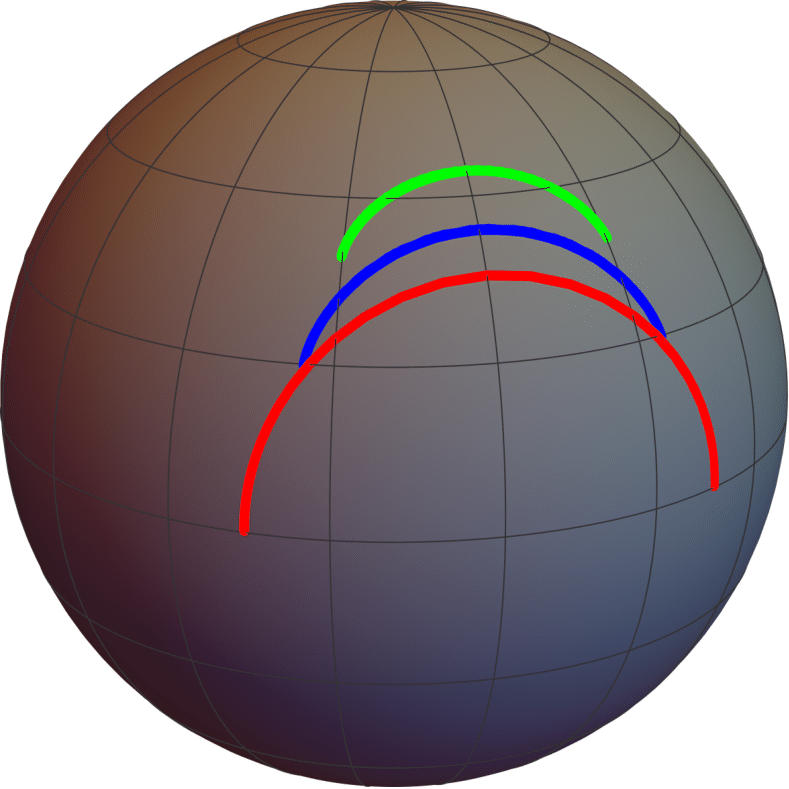}
			\caption{ Giant Magnon}
			\label{fig:sub1}
		\end{subfigure}%
		\begin{subfigure}{.5\textwidth}
			\centering
			\includegraphics[width=.7\linewidth]{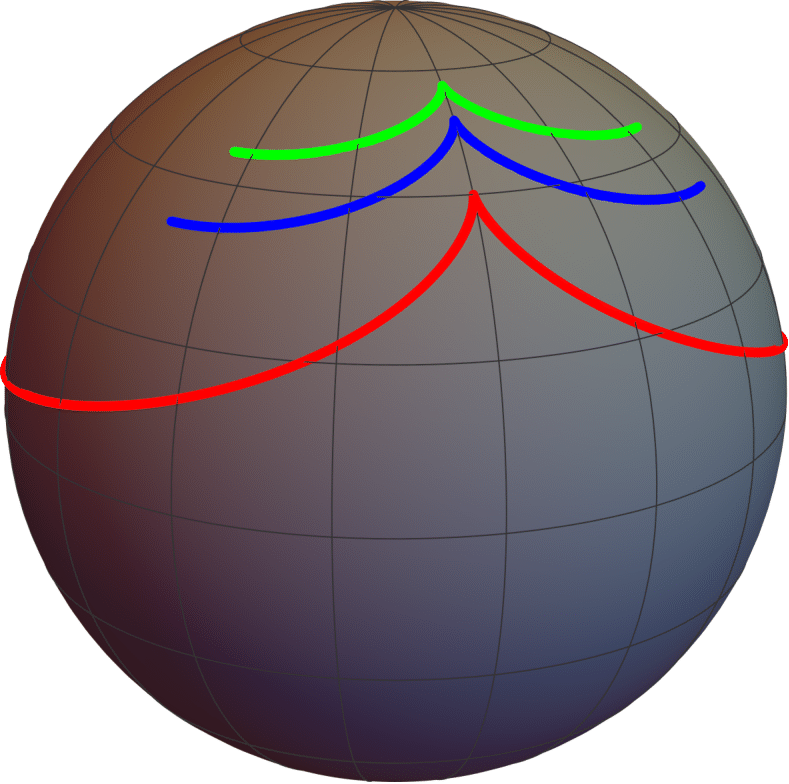}
			\caption{Single spike }
			\label{fig:sub2}
		\end{subfigure}
		\caption{Infinite(red) and finite (blue and green) magnon and single spike for $\varkappa=0.5$ }
		\label{fig:giant magnon and spiky string}
	\end{figure}

For $\alpha \geq 1 $ and $\beta \geq 1 $ with $\frac{\beta}{\alpha} \leq 1$ one can verify that
	\begin{equation}
	z_{max}=\cos^2\theta_0 \quad \text{and} \quad z_{min}=\cos^2\theta_1 \ .
	\end{equation}
	Now integrating out (\ref{phieqn}) and (\ref{thetaeqn}), we find the following solution for
	$\theta$ and $\phi$ respectively 
	\begin{equation}
	z=\cos\theta=z_{max}~ \text{dn}\left[\frac{(\sigma- \beta \tau) B z_{max}}{1-\beta^2},1-\varepsilon\right] \ , 
	\end{equation}
	 \begin{multline}
	\varphi = \alpha \tau - \frac{ \beta(\alpha^2+\varkappa^2)}{\sqrt{(\alpha^2-1)(\alpha^2+\varkappa^2 \beta^2)}}\left(\mathbf{F} \left[\sin^{-1}\left(\frac{1}{\sqrt{1-\varepsilon}} \sqrt{1-\frac{z^2}{z_{max}^2}}\right),1-\varepsilon\right] \right. \\ \left.-  \mathbf{\Pi}\left[  \frac{z_{min}^2-z_{max}^2}{1- z_{max}^2},\sin^{-1}\left(\frac{1}{\sqrt{1-\varepsilon}} \sqrt{1-\frac{z^2}{z_{max}^2}}\right),1-\varepsilon\right]\right).	 
	\end{multline}
	The above solutions are plotted over the deformed $S^2$ sphere  with different value of $\alpha$ and fixed $\beta=1.5$ as shown in the figure:\ref{fig:giant magnon and spiky string}. When $\alpha=\beta$ the above solution reduces to the infinite single spike solution.
	 The conserved charges and the  angular deficit turn out to be
	\begin{equation}
	E=\frac{\sqrt{\lambda(1+\varkappa^2)}}{\pi}\frac{(1-\varepsilon
		)z_{max}}{\sqrt{1-z_{max}^2}\sqrt{1+\varepsilon\varkappa^2z_{max}^2}}\mathbf{K}(1-\varepsilon) \ ,
	\end{equation}
	\begin{equation}
	J=\frac{\sqrt{\lambda(1+\varkappa^2)}}{\pi}\frac{\left( (1+\varkappa^2z_{max}^2)\mathbf{K}(1-\varepsilon)-(1+\varepsilon\varkappa^2 z_{max}^2)\mathbf{\Pi}\left[\frac{(1-\varepsilon)\varkappa^2z_{max}^2}{1+\varkappa^2z_{max}^2},1-\varepsilon\right]\right)}{\varkappa^2z_{max}\sqrt{1+\varkappa^2 z_{max}^2}\sqrt{1+\varepsilon \varkappa^2 z_{max}^2}} \ ,
	\end{equation}
	\begin{equation}
	\Delta \phi= p = \frac{2\sqrt{1-\varepsilon z_{max}^2}\left(\mathbf{K}(1-\varepsilon)-\mathbf{\Pi}\left[\frac{z_{max}^2(1-\varepsilon)}{z_{max}^2-1},1-\varepsilon\right]\right)}{z_{max}\sqrt{1-z_{max}^2}} \ .
	\end{equation}
	Inverting the linear momentum we get $z_{max}$ as
	\begin{equation}\label{singlespike z inverse}
	z_{max}=\sin \xi+ \frac{\left(p\cos \xi+2 \xi\cos \xi+\ sin \xi \right)(1+\varkappa^2 \sin^2 \xi)\varepsilon}{4}+\mathcal{O}(\varepsilon^2)
	\end{equation}
	plugging $z_{max}$ expansion in angular momentum and then inverting it for $\varepsilon$, we get
	
	\begin{equation} \label{singlespike momentum inverse}
	\varepsilon= 16~\exp\left(-(p+2\xi)\cot\xi\right)  \ ,  
	\end{equation}
where $$\sin\xi=\frac{\sinh\left[\frac{J \varkappa}{\sqrt{1+\varkappa^2}}\right]}{\varkappa}.$$
Now substituting (\ref{singlespike momentum inverse}) and (\ref{singlespike z inverse}) in the expansion of $\text{E}-\hat{T} \Delta \phi$, we get the leading order finite size scaling relation for single spike
	\begin{equation}
	\text{E}-\hat{T} \Delta \phi=2 \hat{T}\left(\xi+ \frac{ (1+\varkappa^2)\sin^2 \xi \tan\xi }{4}~ \varepsilon\right).
	\end{equation}
	For $p \rightarrow \infty$ we find
	\begin{equation}
	\text{E}-\hat{T} \Delta \phi=2 \hat{T} \sin^{-1} \left(\frac{\sinh\left[\frac{J \varkappa}{\sqrt{1+\varkappa^2}}\right]}{\varkappa}\right),
	\end{equation}
	which matches with the infinite single spike scaling relation \cite{Banerjee:2014bca}. Similarly taking $\varkappa \rightarrow 0$, the above result reproduces the finite size  scaling relation for single spike in undeformed R$\times $S$^2$ \cite{Ahn:2008sk}. 
%
%

	\section{Spinning folded string solution  }
\begin{figure}[H]
	\centering
	\begin{subfigure}{.5\textwidth}
		\centering
		\includegraphics[width=.8\linewidth]{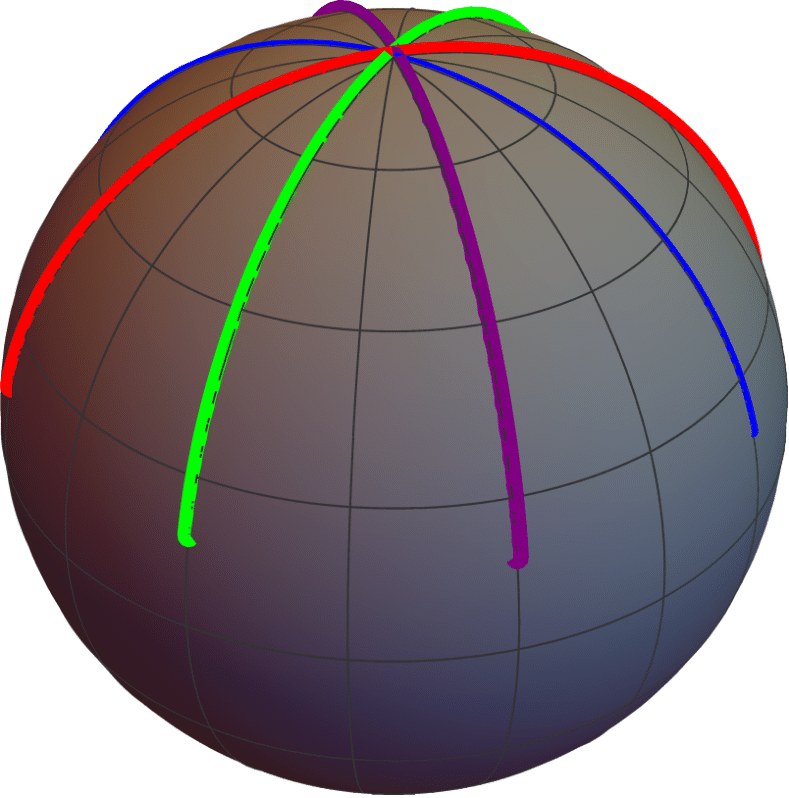}
		\label{fig:sub1}
	\end{subfigure}%
	\begin{subfigure}{.5\textwidth}
		\centering
		\includegraphics[width=.8\linewidth]{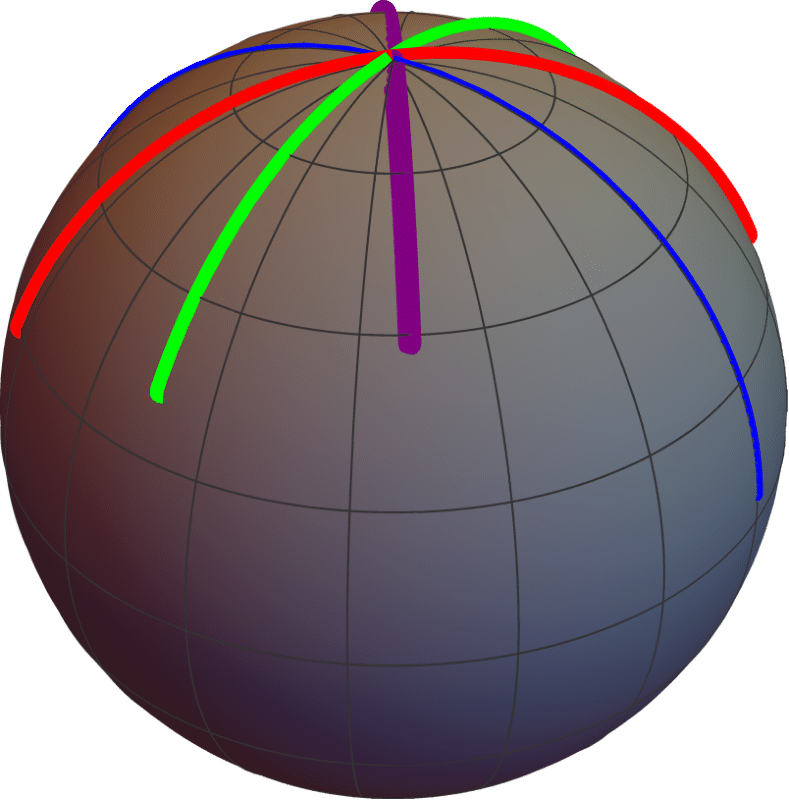}
		\label{fig:gkp string}
	\end{subfigure}
	\caption{Left: Time evolution of GKP string solution for  $\varkappa=1.5$ and $\omega=1.1$:  blue($\tau=0$), red($\tau=\frac{T}{4}$), green($\tau=\frac{3T}{2}$), purple($\tau=\frac{T}{2}$). Right: GKP string solution with different value of $\omega$ but fixed $\varkappa=1.5$.}
	\label{fig:GKP string}
\end{figure}

\begin{figure}[H]
	\centering
	\begin{subfigure}{.5\textwidth}
		\centering
		\includegraphics[width=1\linewidth]{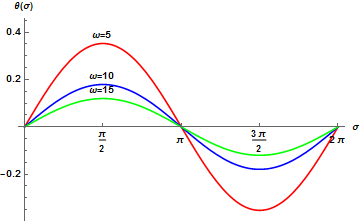}
		\label{fig:sub1}
	\end{subfigure}%
	\begin{subfigure}{.5\textwidth}
		\centering
		\includegraphics[width=1\linewidth]{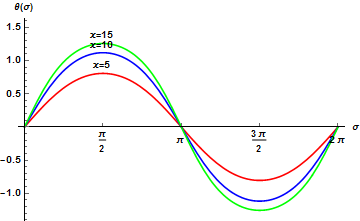}
		\label{}
	\end{subfigure}
	\caption{Plot of $\theta(\sigma)$ for $\varkappa$=const.(left) and $\omega$=const.(right).}
	\label{fig:gkp periodicity}
\end{figure}

In this section, we discuss spinning folded/GKP string solution in R$ \times $S$^2$ subspace of  $\varkappa$-deformed background. This type of string is closed, folded which rigidly rotates around the north pole of two sphere $S^2$ as depicted in \ref{fig:GKP string}. It is dual to the superposition of two magnon excitations each having maximum momentum $\pi$ in $\mathcal{N}=4$ SYM theory. We derive the dispersion relation for both  infinite and finite angular momentum. The  relevant ansatz  for this type of string is given by
\begin{equation}
	t= k \tau, \quad  \theta(\sigma)=\theta(\sigma+2 \pi), \quad \phi=k \omega \tau.
\end{equation}
The equation of motion of $\theta$ and one of the Virasoro constraint respectively gives

\begin{equation}\label{GKP string theta eom}
\theta''(1+\varkappa^2 \cos^2 \theta)+ \varkappa^2 \sin( 2 \theta) \theta'^2+ k^2\omega^2 \sin  \theta \cos \theta (1+\varkappa^2 \cos^2 \theta)=0.
\end{equation}
\begin{equation}
\theta'^2-k^2 (1+\varkappa^2 \cos^2 \theta)+k^2 \omega^2 \sin^2 \theta=0,
\end{equation}
It can be seen that the equations of motion is consistent with the Virasoro constraint equation. Now integrating the equation of motion we find the following periodic solution of $\theta$. 
\begin{equation}
\sin \theta = 
\sqrt{\frac{1+\varkappa^2}{\varkappa^2+\omega^2}}~\text{sn} \left[k \sqrt{\varkappa^2+\omega^2} \sigma, \frac{1 +\varkappa^2}{1+ \omega^2}\right].
\end{equation} 
Time evolution of above solution has been plotted over deformed $S^2$ sphere in fig:\ref{fig:GKP string}. For $\omega=1$ it corresponds to infinite size spinning folded string solution.
From (\ref{GKP string theta eom}) we find the the maximum value  of $\theta$ is
\begin{equation}
	\theta_{max}= \sin^{-1} \sqrt{\frac{1+\varkappa^2}{\varkappa^2+\omega^2}}
\end{equation}
The periodicity condition of $\theta$ implies, it increases for $0\leq \sigma\leq \frac{\pi}{2}$, and reaches $\theta_{max}$ at $\sigma= \frac{\pi}{2}$, then it decreases to zero for $\frac{\pi}{2} \leq \sigma \leq \pi$ and continues the pattern for the other half as shown in fig:\ref{fig:gkp periodicity} for different values of $\varkappa$ and $\omega$.
\begin{equation}
2\pi= \int_0^ {2\pi} d\sigma = 4 \int_0^ {\theta_{max}} \frac{d \theta}{k \sqrt{1+\varkappa^2}\sqrt{1- \frac{\varkappa^2+\omega^2}{1+\varkappa^2} \sin^2 \theta}}
\end{equation}
\begin{equation}
\text{or}~~~~~	\frac{\pi}{2} k \sqrt{\varkappa^2+\omega^2}= \mathbf{K} \left(\frac{1+\varkappa^2}{\varkappa^2+\omega^2}\right)
\end{equation}
 The conserved charges and the difference between them can be calculated as 
 \begin{equation}
 	\text{E}= \frac{\hat{T}}{2}
 \int \frac{\partial \mathcal{L}}{\partial \dot{t}}d \sigma =\frac{4 \hat{T}}{\sqrt{1+\varkappa^2}} \int_{0}^{\theta_{max}} \frac{d \theta}{\sqrt{1- \frac{\varkappa^2+ \omega^2}{1+\varkappa^2} \sin^2 \theta}}= \frac{2 \sqrt{\lambda}}{\pi} ~\mathbb{F} \left[\theta_{max}, \frac{\varkappa^2+\omega^2}{1+\varkappa^2}\right], 
 \end{equation}
 \begin{multline}
 	\text{J}=\frac{\hat{T}}{2}
 	\int \frac{\partial \mathcal{L}}{\partial \dot{\phi}}d \sigma=\frac{4 \hat{T}}{\sqrt{1+\varkappa^2}} \int_{0}^{\theta_{max}} \frac{\omega \sin^2 \theta}{1+ \varkappa^2 \cos^2 \theta} \frac{d \theta}{\sqrt{1- \frac{ \varkappa^2 +\omega^2}{1+\varkappa^2} \sin^2 \theta}}\\= \frac{2 \sqrt{\lambda} \omega}{\pi \varkappa^2} \left( \mathbf{\Pi}\left[\frac{\varkappa^2}{1+\varkappa^2}, \theta_{max},\frac{\varkappa^2+\omega^2}{1+\varkappa^2} \right]-\mathbb{F} \left[\theta_{max}, \frac{\varkappa^2+\omega^2}{1+\varkappa^2}\right]\right).
 \end{multline}
   \begin{equation}
 \text{E}-\text{J}= \frac{4 \hat{T}}{\sqrt{1+\varkappa^2}} \int_{0}^{\theta_{max}} \frac{d \theta}{\sqrt{1- \frac{\varkappa^2+\omega^2}{1+\varkappa^2} \sin^2 \theta} } \left(1- \frac{\omega \sin^2 \theta}{1+\varkappa^2 \cos^2 \theta}\right)
 \end{equation}
 \begin{figure}[!htb]
 	\centering
 	\begin{subfigure}{.5\textwidth}
 		\centering
 		\includegraphics[width=1\linewidth]{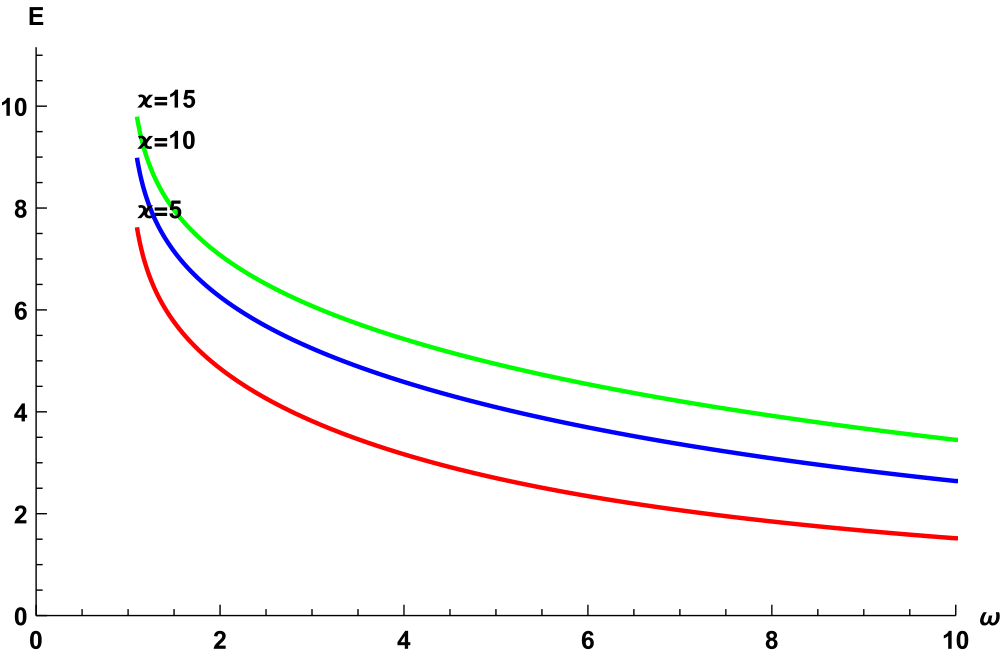}
 		\label{fig:sub1}
 	\end{subfigure}%
 	\begin{subfigure}{.5\textwidth}
 		\centering
 		\includegraphics[width=1\linewidth]{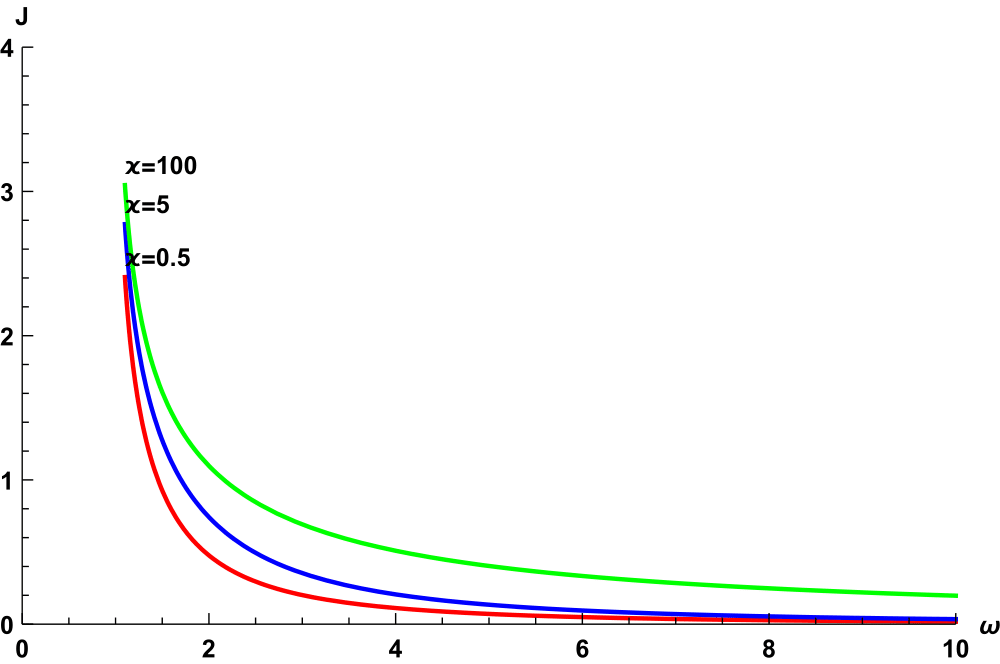}
 		\label{fig:sub2}
 	\end{subfigure}
 	\caption{Plot for energy(left) and spin (right) for different values of $\varkappa$}
 	\label{fig:gkp energy and angular momentum}
 \end{figure}

  Depending upon the value of the $\frac{\varkappa^2+\omega^2}{1+\varkappa^2}$ we have two different kind of string solution  (i) For $\frac{\varkappa^2+\omega^2}{1+\varkappa^2}>1 $: Folded  closed string  solution. (ii) $\frac{\varkappa^2+\omega^2}{1+\varkappa^2}<1 $ circular string solution. Here we will discuss about only folded string solution for which the conserved charges and their different turn out to be 
  \begin{eqnarray}
  	\text{E}&=& \frac{4\hat{T}}{\sqrt{\varkappa^2+\omega^2}} \mathbf{K} \left[\frac{1+\varkappa^2}{\varkappa^2+\omega^2}\right], \\
  	\text{J}&=& \frac{4\hat{T}}{\varkappa^2\sqrt{\varkappa^2+\omega^2}} \left(\mathbf{\Pi}\left[\frac{\varkappa^2}{\varkappa^2+\omega^2}, \frac{1+\varkappa^2}{\varkappa^2+\omega^2}\right]-\mathbf{K} \left[\frac{1+\varkappa^2}{\varkappa^2+\omega^2}\right]\right), \\
  	\text{E}-\text{J}&=&\frac{4\hat{T}}{\sqrt{\varkappa^2+\omega^2}}\left( (\varkappa^2-1)\mathbf{K} \left[\frac{1+\varkappa^2}{\varkappa^2+\omega^2}\right] -\mathbf{\Pi}\left[\frac{\varkappa^2}{\varkappa^2+\omega^2}, \frac{1+\varkappa^2}{\varkappa^2+\omega^2}\right] \right). \label{gkp e-j}
  \end{eqnarray}
   Here we can see when the string is stretched to the equator i.e. $\theta_{max}= \frac{\pi}{2}$ or $\omega=1$, both E and J diverge (fig:\ref{fig:gkp energy and angular momentum}) while their difference remains finite

   \begin{equation}\label{GKP string infinite charge }
   	\text{E}-\text{J}=\frac{2 \sqrt{\lambda}}{\pi} \int_{0}^{\frac{\pi}{2}} \frac{  (1+\varkappa^2)\cos \theta d \theta}{1+\varkappa^2 \cos^2 \theta}= \frac{2 \sqrt{\lambda}}{\pi} \frac{\sinh^{-1}\varkappa}{\varkappa}
   \end{equation}
   This is the scaling relation of GKP string in (R$\times$ S$^2$)$_\varkappa$  with infinite energy and angular momentum. For $\varkappa \rightarrow 0$ the above result reduces to 
   \begin{equation}
   \text{E}-\text{J}=\frac{2 \sqrt{\lambda}}{\pi},
   \end{equation}
   which matches with the dispersion relation for spinning folded string solution in undeformed R $\times$ S$^2$ as found in \cite{Gubser:2002tv}. It can also be verified that, the above relation matches with the scaling relation of the superposition of two giant magnon when they both have momentum $p=\pi$. Now we want to find the scaling relation for finite angular momentum by expanding both energy and angular momentum around $\theta_{max}=\frac{\pi}{2}$ or $\omega=1$. Taking $\varepsilon=1-\frac{1+\varkappa^2}{\varkappa^2+\omega^2}$ we can rewrite the conserved charges as
  \begin{equation}
  	\text{E}= \frac{2\sqrt{\lambda}}{\pi} \sqrt{1-\varepsilon}~ \mathbf{K} \left[1-\varepsilon\right],
  \end{equation}
  \begin{equation}
  	\text{J}=  \frac{2\sqrt{\lambda}\sqrt{1-\varepsilon}}{\pi \varkappa^2} ~ \left(\mathbf{\Pi}\left[\frac{\varkappa^2}{1+\varkappa^2}, 1-\varepsilon\right]-\mathbf{K} \left[1-\varepsilon\right]\right).
  \end{equation} 
  Now shifting the argument of the Jacobi elliptic function we can isolate the singularity, for which the angular momentum becomes
  \begin{equation}
  \mathcal{J}= \frac{\pi\text{J}}{\sqrt{\lambda}}  = i \pi \sqrt{1+\varkappa^2} \frac{\mathbf{F}\left[ \sin^{-1} i \varkappa, \varepsilon\right]}{\mathbf{K}(\varepsilon)}+ \frac{2(1-\varepsilon)}{\sqrt{1+\varepsilon \varkappa^2}}\frac{\mathbf{K}(1-\varepsilon) \mathbf{\Pi}\left[\frac{\varepsilon(1+\varkappa^2)}{1+\varepsilon \varkappa^2}, \varepsilon\right]}{~\mathbf{K}\left(\varepsilon\right)}.
  \end{equation}
  Now we expand  $\mathcal{J}$ around $\varepsilon=0$ and using the expansion of elliptic function, we get $\mathcal{J}$ in following form 
  \begin{equation}
  \mathcal{J}=\sum_{n=0}^\infty \varepsilon^n(c_n \ln \varepsilon+ b_n) \Rightarrow \ln \varepsilon = \frac{\mathcal{J}- \sum_{n=0}^\infty b_n \varepsilon^n}{\sum_{n=0} c_n \varepsilon^n}.
  \end{equation}
  After exponentiating and rearranging it, we get
  
  \begin{equation}\label{e0expansion}
  \varepsilon_0=  \varepsilon \exp \left(\sum_{n=0}^\infty b_n \varepsilon^n -\left(\frac{\mathcal{J}-b_0}{c_0}-\sum_{n=0}^\infty b_n \varepsilon^n\right)\sum_{n=1}^\infty (-1)^n \left(\sum_{k=1}^\infty \frac{c_k}{c_0}\varepsilon^k\right)^n\right)
  \end{equation}
  
   where $$\varepsilon_0= \exp \left(\frac{\mathcal{J}-b_0}{c_0}\right)=16 ~e^{-\mathcal{J}-\frac{2 \sqrt{1+\varkappa^2} \sinh^{-1}{ \varkappa}}{\varkappa}} $$ is the lowest order in  $\varepsilon$ . The above expansion can be written in the following form 
  $\varepsilon_0= \sum_{n=1}^{\infty} a_n \varepsilon^n$ where the coefficient $a_n$ can be found from (\ref{e0expansion}). Now we can use Lagrange inverse formula to find $\varepsilon$ in the series expansion of $\varepsilon_0$. We find the inverse series in the following form
   \begin{equation}\label{GKPinverseJform}
   \varepsilon = \sum_{n=1}^\infty \varepsilon_0^n \sum_{m=0}^{n-1} c_{nm} \mathcal{J}^m
   \end{equation} the coefficients $c_{nm}$ is independent of $\mathcal{J}$. We find the following  leading order terms 
   \begin{eqnarray}\label{gkp leading order e}
    \varepsilon_{leading}&=& \sum_{n=1}^\infty  c_{n,n-1} \mathcal{J}^{n-1} \varepsilon_0^n \nonumber \\
  & =& \varepsilon_0-\frac{\mathcal{J} \varepsilon_0^2}{4}+\frac{ 3 \mathcal{J}^2 \varepsilon_0^3}{32}-\frac{1}{24} \mathcal{J}^3  \varepsilon_0^4+\frac{125}{6144} \mathcal{J}^4 \varepsilon_0^5-\frac{27}{2560} \mathcal{J}^5 \varepsilon_0^6+\cdots . \nonumber \\
   \end{eqnarray}
   The above leading order terms of $\varepsilon$ can be generalized and subsequently can be written in terms of Lambert's W function as
   \begin{equation}
   \varepsilon_{leading}= \frac{4}{\mathcal{J}} \sum_{n=1}^\infty \frac{(-n)^{n-1}}{n!} \frac{\varepsilon_0 \mathcal{J}}{4}=\frac{4}{\mathcal{J}} W \left( \frac{\varepsilon_0\mathcal{J}}{4}\right).
   \end{equation}
   Finally  substituting (\ref{gkp leading order e}) in  (\ref{gkp e-j}), we get the following dispersion relation up to leading order as
     \begin{multline}\label{gkpfinite}
   \mathcal{E}-\mathcal{J}|_{leading}=  \frac{2\sqrt{1+\varkappa^2} \sinh^{-1} \varkappa}{\varkappa}-\frac{(1+\varkappa^2)\varepsilon_0}{2}\left(1-\frac{\mathcal{J}\varepsilon_0}{8}+\frac{\mathcal{J}^2\varepsilon_0^2}{32}+\cdots\right). 
 \end{multline}
 The above relation can be written in compact form using Lambert W function as
    
  \begin{equation}
  \mathcal{E}-\mathcal{J}|_{leading}=\frac{2\sqrt{1+\varkappa^2} \sinh^{-1} \varkappa}{\varkappa}-\frac{1+\varkappa^2}{\mathcal{J}}(2 W+W^2).
  \end{equation}
    We can see for $\mathcal{J}\rightarrow \infty$, (\ref{gkpfinite}) gives
    \begin{equation}
    \mathcal{E}-\mathcal{J}=2 \frac{\sinh^{-1}\varkappa}{\varkappa}.
    \end{equation}
   which is the infinite size scaling relation for folded spinning string as derived  in (\ref{GKP string infinite charge }). Similarly for $\varkappa \rightarrow 0$ it gives result obtained in \cite{Arutyunov:2006gs} for undeformed $R_t \times S^2$.
   \section{Concluding remarks}
   In this paper, we have found the finite size scaling relation for various classical string solutions in $\varkappa$-deformed $AdS_3\times S^3$ by expanding the conserved charges at their large values. First we have computed the finite size correction to the scaling relation of giant magnon and single spike string solutions in $\varkappa$- deformed $R_t \times S^2$ geometry.  For the giant magnon case the expression for the finite size correction matches with the L\"{u}scher correction obtained from the exact S-matrices at strong coupling. For the  single spike string solution one needs to find the corrections at strong coupling and compare it with our present result. We have shown for $\mathcal{J}\rightarrow \infty$ in giant magnon case and $p\rightarrow \infty$ in single spiky case, the finite size correction reproduce the corresponding infinite size scaling relation. Further we have also obtained the finite size correction to folded spinning string solution. For $\varkappa \rightarrow 0$ all our results reduce to the results obtained in undeformed $R_t \times$S$^2$. \par
   
 The results obtained here can be extended in a number of directions. First, it will be interesting to find out the finite size expansion for magnon bound state or dyonic giant magnon in $\varkappa$-deformed background where correction terms will depend on two angular momenta along the $S^3$. It will also be worth studying similar finite size effect for the classical string solutions in AdS$_3 \times$ S$^3$ with mixed flux. We wish to report on these issues in near future.
  \section*{Acknowledgements}
   We would like to thank G. Linardopoulos for some useful discussions. 
   \appendix \section{Infinite charge giant magnon}
  
   Here we review the dispersion relation of giant magnon having infinite energy and angular momentum\cite{Arutynov:2014ota,Khouchen:2014kaa,Banerjee:2014bca}. In this case the end points of the string lie on the equator i.e $\theta_0=\frac{\pi}{2}$. For which the conserved charges and the angular difference becomes
   \begin{equation}
   \text{E}= 2 \hat{T} \frac{1- \beta^2}{B} \int_{\theta_1}^{\frac{\pi}{2}}  \frac{\sin \theta}{\cos \theta\sqrt{\cos^2 \theta_1- \cos^2 \theta}}
   \end{equation}
   \begin{equation}
   \text{J}= 2 \hat{T} \frac{1- \beta^2}{B}  \int_{\theta_1}^{\frac{\pi}{2}}  \frac{\sin \theta}{\cos \theta\sqrt{\cos^2 \theta_1- \cos^2 \theta}}\frac{\beta^2(1+\varkappa^2 \cos^2 \theta)-\sin^2 \theta}{(\beta^2-1)(1+\varkappa^2 \cos^2 \theta)}
   \end{equation}
   We can see that both E and J diverge but their difference remains finite.
   \begin{eqnarray}\label{energy ang diff}
   \text{E}-\text{J}&=& \frac{2 \hat{T}}{\sqrt{(1+\varkappa^2)(1+\varkappa^2 \beta^2)}}\int_{\theta_1}^{\frac{\pi}{2}}\frac{d \theta\sin \theta \cos\theta}{(1+\varkappa^2 \cos \theta)\sqrt{\cos^2\theta_1- \cos^2 \theta}} \nonumber \\
   &=& 2 \hat{T}  \tanh^{-1} \frac{ \varkappa \cos \theta_1}{\sqrt{1+\varkappa^2 \cos^2 \theta_1}}
   \end{eqnarray}
   The angular difference  can be calculated as
   \begin{eqnarray}\label{ang diff}
   \Delta \phi &=& 2 \frac{ \beta (1+\varkappa^2)}{\sqrt{(1+\varkappa^2)(1+\varkappa^2 \beta^2)}} \int_{\theta_1}^{\frac{\pi}{2}}\frac{\sin \theta}{\cos \theta\sqrt{\cos^2 \theta_1- \cos^2 \theta}} \nonumber \\
   &=& 2 \sin^{-1} (\cos \theta_1)
   \end{eqnarray}
    Combining  (\ref{energy ang diff}) and (\ref{ang diff}), we find
    \begin{eqnarray}
    \text{E}-\text{J}&=& \frac{2 \hat{T}}{\varkappa} \tanh^{-1}\frac{\varkappa | \sin \frac{\Delta \phi}{2}|}{\sqrt{1+\varkappa^2 \sin^2 \frac{\Delta \phi}{2}}} \nonumber \\
\text{or} ~~~ \text{E}-\text{J} &=& \frac{2 \hat{T}}{\varkappa} \sinh^{-1}\left(\varkappa \sin \frac{\Delta \phi}{2}\right)
    \end{eqnarray}
    \section{Infinite charge single spike}
    In this case, $ \theta_1 = \frac{\pi}{2}$. For which the energy and angular difference  are infinite where the difference between them and the angular momentum remains finite \cite{Banerjee:2014bca}.
    \begin{equation}
    \text{E}=\frac{2\hat{T}(1-\alpha^2) }{\sqrt{(\alpha^2+\varkappa^2)(1+\varkappa^2)}}\int_{\frac{\pi}{2}}^{\theta_0}  \frac{d \theta\sin \theta}{\cos \theta\sqrt{\cos^2 \theta_0- \cos^2 \theta}}
    \end{equation}
    \begin{equation}
    \Delta \phi=\frac{2(1-\alpha^2) }{\sqrt{(\alpha^2+\varkappa^2)(1+\varkappa^2)}}\int_{\frac{\pi}{2}}^{\theta_0}  \frac{d \theta\sin \theta}{\cos \theta\sqrt{\cos^2 \theta_0- \cos^2 \theta}}\frac{\alpha^2- \frac{1+ \varkappa^2 \cos^2 \theta}{\sin^2 \theta}}{a^2-1}
    \end{equation}
    \begin{eqnarray}
    \text{E}- \hat{T} \Delta \phi & =& -\frac{2 \hat{T} \sqrt{1+\varkappa^2}}{\sqrt{\alpha^2+\varkappa^2}}\int_{\frac{\pi}{2}}^{\theta_0}\frac{\cos \theta}{\sin \theta\sqrt{\cos^2 \theta_0- \cos^2 \theta}} \nonumber \\
    &=&  2 \hat{T}\sin^{-1} \cos \theta_0 \nonumber \\
    &=& 2 \hat{T} \left(\frac{\pi}{2}- \theta_0\right) \label{infinte spiky E-phi }
    \end{eqnarray}
    The angular momentum can be calculated as
    \begin{eqnarray}
    \text{J} &=&\frac{-2\hat{T}(1+\varkappa^2) }{\sqrt{(\alpha^2+\varkappa^2)}}\int_{\frac{\pi}{2}}^{\theta_0}  \frac{d \theta\sin \theta \cos \theta}{(1+\varkappa^2 \cos^2 \theta)\sqrt{\cos^2 \theta_0- \cos^2 \theta}} \nonumber \\
    &=& \frac{2 \hat{T}}{\varkappa} \tanh^{-1} \frac{\varkappa \cos \theta_0}{\sqrt{1+\varkappa^2 \cos^2 \theta_0}} \label{infinite spiky j}
    \end{eqnarray}
    combining (\ref{infinte spiky E-phi }) and (\ref{infinite spiky j}), gives
    \begin{equation}
    \text{E}- \hat{T} \Delta \phi=2 \hat{T}\sin^{-1} \left(\frac{\sinh\left[\frac{J \varkappa}{\sqrt{1+\varkappa^2}}\right]}{\varkappa}\right)
    \end{equation}
     
    \section{Useful formulas of elliptic integrals }
   \begin{eqnarray}
    \int_{z_{min}}^{z_{max}}\frac{dz ~ z_{max}}{\sqrt{(z_{max}^2-z_{min}^2)(z^2-z_{min}^2)}}&=&\mathbf{K}\left(1-\frac{z_{min}^2}{z_{max}^2}\right) \nonumber \\
    \int_{z_{min}}^{z_{max}}\frac{dz ~ z^2}{z_{max}\sqrt{(z_{max}^2-z_{min}^2)(z^2-z_{min}^2)}}&=&\mathbf{K}\left(1-\frac{z_{min}^2}{z_{max}^2}\right) \nonumber \\
     \int_{z_{min}}^{z_{max}}\frac{dz ~ z_{max}(1-z_{max}^2)}{(1-z^2)\sqrt{(z_{max}^2-z_{min}^2)(z^2-z_{min}^2)}} &=& \mathbf{\Pi}\left(\frac{z_{max}^2-z_{min}^2}{z_{max}^2-1},\frac{z_{min}^2}{z_{max}^2}\right)
   \end{eqnarray}
   Modulus transformation identities
   \begin{eqnarray}
   \mathbf{F} \left[\phi,m\right]=\frac{1}{\sqrt{m}} \mathbf{F}\left[\phi_1,\frac{1}{m}\right] \nonumber \\
   \mathbf{\Pi} \left[n,\phi,m\right]= \frac{1}{\sqrt{m}} \mathbf{\Pi}\left[\frac{n}{m}, \phi_1, \frac{1}{m}\right]
   \end{eqnarray}
   where  $\phi_1= \sin^{-1}(\sqrt{m} \sin \phi)$.
    \begin{eqnarray}
   \mathbf{F} \left[\phi,m\right]= -i \mathbf{F} \left[\sec^{-1} (\cos \phi),1-m\right] \nonumber \\
   \mathbf{\Pi} \left[n,\phi,m\right]= \frac{i}{1-n}\left(n  \mathbf{\Pi} \left[1-n,\sec^{-1} (\cos \phi),1-m \right]-\mathbf{F}\left[\sec^{-1} (\cos \phi),1-m\right]\right) \nonumber \\
   \end{eqnarray}
   Addition Formula
   \begin{multline}\label{addition formula}
   \mathbf{\Pi}\left[n,m\right]=\frac{1}{(1-n) \mathbf{K}(1-m)} \left(\frac{\pi}{2} \sqrt{\frac{n(n-1)}{m-n}} \mathbf{F}\left[\sin^{-1}\sqrt{\frac{n}{n-m}},1-m\right]- \right.\\  \left.\mathbf{K}(m)\left[(n-1) \mathbf{K}(1-m)-n\mathbf{\Pi}\left[\frac{1-m}{1-n},1-m\right]\right]\right)
   \end{multline}


\begin{thebibliography}{99}
    	\bibitem{Maldacena:1997re} 
    	J.~M.~Maldacena,
    	``The Large N limit of superconformal field theories and supergravity,''
    	Int.\ J.\ Theor.\ Phys.\  {\bf 38}, 1113 (1999)
    	[Adv.\ Theor.\ Math.\ Phys.\  {\bf 2}, 231 (1998)]
    	[hep-th/9711200].
    	
    	
    \bibitem{Witten:1998qj} 
    E.~Witten,
    ``Anti-de Sitter space and holography,''
    Adv.\ Theor.\ Math.\ Phys.\  {\bf 2}, 253 (1998)
    [hep-th/9802150].
    
    \bibitem{Gubser:1998bc} 
    S.~S.~Gubser, I.~R.~Klebanov and A.~M.~Polyakov,
    ``Gauge theory correlators from noncritical string theory,''
    Phys.\ Lett.\ B {\bf 428}, 105 (1998)
    [hep-th/9802109].
    
    
    \bibitem{Minahan:2002ve} 
    J.~A.~Minahan and K.~Zarembo,
    ``The Bethe ansatz for N=4 superYang-Mills,''
    JHEP {\bf 0303}, 013 (2003)
    doi:10.1088/1126-6708/2003/03/013
    [hep-th/0212208].
    
    
    \bibitem{Beisert:2003yb} 
    N.~Beisert and M.~Staudacher,
    ``The N=4 SYM integrable super spin chain,''
    Nucl.\ Phys.\ B {\bf 670}, 439 (2003)
    [hep-th/0307042].
    
    \bibitem{Beisert:2004hm} 
    N.~Beisert, V.~Dippel and M.~Staudacher,
    ``A Novel long range spin chain and planar N=4 super Yang-Mills,''
    JHEP {\bf 0407}, 075 (2004)
    [hep-th/0405001].
    
    \bibitem{Beisert:2005fw} 
    N.~Beisert and M.~Staudacher,
    ``Long-range psu(2,$2|4$) Bethe Ansatze for gauge theory and strings,''
    Nucl.\ Phys.\ B {\bf 727}, 1 (2005)
    [hep-th/0504190].
    
    \bibitem{Metsaev:1998it} 
    R.~R.~Metsaev and A.~A.~Tseytlin,
    ``Type IIB superstring action in AdS(5) x S**5 background,''
    Nucl.\ Phys.\ B {\bf 533}, 109 (1998)
    [hep-th/9805028].
    
    \bibitem{Bena:2003wd} 
    I.~Bena, J.~Polchinski and R.~Roiban,
    ``Hidden symmetries of the AdS(5) x S**5 superstring,''
    Phys.\ Rev.\ D {\bf 69}, 046002 (2004)
    [hep-th/0305116].
    
    
    \bibitem{Gubser:2002tv} 
    S.~S.~Gubser, I.~R.~Klebanov and A.~M.~Polyakov,
    ``A Semiclassical limit of the gauge / string correspondence,''
    Nucl.\ Phys.\ B {\bf 636}, 99 (2002)
    doi:10.1016/S0550-3213(02)00373-5
    [hep-th/0204051].
    
 
    
    
    \bibitem{Minahan:2002rc} 
    J.~A.~Minahan,
    ``Circular semiclassical string solutions on AdS(5) x S(5),''
    Nucl.\ Phys.\ B {\bf 648}, 203 (2003)
    [hep-th/0209047].
    
    \bibitem{Tseytlin:2004xa} 
    A.~A.~Tseytlin,
    ``Semiclassical strings and AdS/CFT,''
    hep-th/0409296.
    
    
\bibitem{Hofman:2006xt} 
D.~M.~Hofman and J.~M.~Maldacena,
``Giant Magnons,''
J.\ Phys.\ A {\bf 39}, 13095 (2006)
[hep-th/0604135].

\bibitem{Dorey:2006dq} 
N.~Dorey,
``Magnon Bound States and the AdS/CFT Correspondence,''
J.\ Phys.\ A {\bf 39}, 13119 (2006)
[hep-th/0604175].

\bibitem{Chen:2006gea} 
H.~Y.~Chen, N.~Dorey and K.~Okamura,
``Dyonic giant magnons,''
JHEP {\bf 0609}, 024 (2006)
[hep-th/0605155].

\bibitem{Kruczenski:2006pk} 
M.~Kruczenski, J.~Russo and A.~A.~Tseytlin,
``Spiky strings and giant magnons on S**5,''
JHEP {\bf 0610}, 002 (2006)
doi:10.1088/1126-6708/2006/10/002
[hep-th/0607044].

\bibitem{Kruczenski:2004wg} 
M.~Kruczenski,
``Spiky strings and single trace operators in gauge theories,''
JHEP {\bf 0508}, 014 (2005)
[hep-th/0410226].

\bibitem{Ishizeki:2007we} 
R.~Ishizeki and M.~Kruczenski,
``Single spike solutions for strings on S**2 and S**3,''
Phys.\ Rev.\ D {\bf 76}, 126006 (2007)
[arXiv:0705.2429 [hep-th]].



\bibitem{Tseytlin:2003ii} 
A.~A.~Tseytlin,
``Spinning strings and AdS / CFT duality,''
In *Shifman, M. (ed.) et al.: From fields to strings, vol. 2* 1648-1707
[hep-th/0311139].

\bibitem{Arutyunov:2003uj} 
G.~Arutyunov, S.~Frolov, J.~Russo and A.~A.~Tseytlin,
``Spinning strings in AdS(5) x S**5 and integrable systems,''
Nucl.\ Phys.\ B {\bf 671}, 3 (2003)
[hep-th/0307191].

\bibitem{Sieg:2005kd} 
C.~Sieg and A.~Torrielli,
``Wrapping interactions and the genus expansion of the 2-point function of composite operators,''
Nucl.\ Phys.\ B {\bf 723}, 3 (2005)
[hep-th/0505071].

\bibitem{Ambjorn:2005wa} 
J.~Ambjorn, R.~A.~Janik and C.~Kristjansen,
``Wrapping interactions and a new source of corrections to the spin-chain/string duality,''
Nucl.\ Phys.\ B {\bf 736}, 288 (2006)
[hep-th/0510171].

\bibitem{Kotikov:2007cy} 
A.~V.~Kotikov, L.~N.~Lipatov, A.~Rej, M.~Staudacher and V.~N.~Velizhanin,
``Dressing and wrapping,''
J.\ Stat.\ Mech.\  {\bf 0710}, P10003 (2007)
[arXiv:0704.3586 [hep-th]].

\bibitem{Arutyunov:2006gs} 
G.~Arutyunov, S.~Frolov and M.~Zamaklar,
``Finite-size Effects from Giant Magnons,''
Nucl.\ Phys.\ B {\bf 778}, 1 (2007)
[hep-th/0606126].


\bibitem{Astolfi:2007uz} 
D.~Astolfi, V.~Forini, G.~Grignani and G.~W.~Semenoff,
``Gauge invariant finite size spectrum of the giant magnon,''
Phys.\ Lett.\ B {\bf 651}, 329 (2007)
[hep-th/0702043 [HEP-TH]].

\bibitem{Minahan:2008re} 
J.~A.~Minahan and O.~Ohlsson Sax,
``Finite size effects for giant magnons on physical strings,''
Nucl.\ Phys.\ B {\bf 801}, 97 (2008)
[arXiv:0801.2064 [hep-th]].

\bibitem{Ramadanovic:2008qd} 
B.~Ramadanovic and G.~W.~Semenoff,
``Finite Size Giant Magnon,''
Phys.\ Rev.\ D {\bf 79}, 126006 (2009)
[arXiv:0803.4028 [hep-th]].

\bibitem{Klose:2008rx} 
T.~Klose and T.~McLoughlin,
``Interacting finite-size magnons,''
J.\ Phys.\ A {\bf 41}, 285401 (2008)
[arXiv:0803.2324 [hep-th]].



\bibitem{Shenderovich:2008bs} 
I.~Shenderovich,
``Giant magnons in AdS(4) / CFT(3): Dispersion, quantization and finite-size corrections,''
arXiv:0807.2861 [hep-th].

\bibitem{Ahn:2008gd} 
C.~Ahn and P.~Bozhilov,
``Finite-size effects of Membranes on AdS(4) x S**7,''
JHEP {\bf 0808}, 054 (2008)
[arXiv:0807.0566 [hep-th]].

\bibitem{Hatsuda:2008na} 
Y.~Hatsuda and R.~Suzuki,
``Finite-Size Effects for Multi-Magnon States,''
JHEP {\bf 0809}, 025 (2008)
[arXiv:0807.0643 [hep-th]].

\bibitem{Ahn:2008sk} 
C.~Ahn and P.~Bozhilov,
``Finite-size Effects for Single Spike,''
JHEP {\bf 0807}, 105 (2008)
doi:10.1088/1126-6708/2008/07/105
[arXiv:0806.1085 [hep-th]].

\bibitem{Bykov:2008bj} 
D.~V.~Bykov and S.~Frolov,
``Giant magnons in TsT-transformed AdS(5) x S**5,''
JHEP {\bf 0807}, 071 (2008)
[arXiv:0805.1070 [hep-th]].

\bibitem{Grignani:2008te} 
G.~Grignani, T.~Harmark, M.~Orselli and G.~W.~Semenoff,
``Finite size Giant Magnons in the string dual of N=6 superconformal Chern-Simons theory,''
JHEP {\bf 0812}, 008 (2008)
[arXiv:0807.0205 [hep-th]].

\bibitem{Bozhilov:2010rf} 
P.~Bozhilov,
``Close to the Giant Magnons,''
arXiv:1010.5465 [hep-th].

\bibitem{Ahn:2010da} 
C.~Ahn and P.~Bozhilov,
``Finite-Size Dyonic Giant Magnons in TsT-transformed $AdS_5\times S^5$,''
JHEP {\bf 1007}, 048 (2010)
[arXiv:1005.2508 [hep-th]].

\bibitem{Jain:2008mt} 
S.~Jain and K.~L.~Panigrahi,
``Spiky Strings in AdS(4) x CP**3 with Neveu-Schwarz Flux,''
JHEP {\bf 0812}, 064 (2008)
[arXiv:0810.3516 [hep-th]].

\bibitem{Floratos:2013cia} 
E.~Floratos, G.~Georgiou and G.~Linardopoulos,
``Large-Spin Expansions of GKP Strings,''
JHEP {\bf 1403}, 018 (2014)
[arXiv:1311.5800 [hep-th]].

\bibitem{Floratos:2014gqa} 
E.~Floratos and G.~Linardopoulos,
``Large-Spin and Large-Winding Expansions of Giant Magnons and Single Spikes,''
Nucl.\ Phys.\ B {\bf 897}, 229 (2015)
[arXiv:1406.0796 [hep-th]].

\bibitem{Ahn:2014aqa} 
C.~Ahn and P.~Bozhilov,
``Finite-size giant magnons on $\eta$-deformed $AdS_5×S^5$,''
Phys.\ Lett.\ B {\bf 737}, 293 (2014)
[arXiv:1406.0628 [hep-th]].

\bibitem{Ahn:2014tua} 
C.~Ahn and P.~Bozhilov,
``String solutions in AdS$_3 \times$ S$^3 \times $ T$^4$ with NS-NS B-field,''
Phys.\ Rev.\ D {\bf 90}, no. 6, 066010 (2014)
[arXiv:1404.7644 [hep-th]].

\bibitem{Ahn:2016egk} 
C.~Ahn,
``Finite-size effect of $\eta$-deformed AdS$_5 \times$ S$^5$ at strong coupling,''
Phys.\ Lett.\ B {\bf 767}, 121 (2017)
[arXiv:1611.09992 [hep-th]].

\bibitem{Delduc:2013qra} 
F.~Delduc, M.~Magro and B.~Vicedo,
``An integrable deformation of the AdS$_5 \times$ S$^5$ superstring action,''
Phys.\ Rev.\ Lett.\  {\bf 112}, no. 5, 051601 (2014)
[arXiv:1309.5850 [hep-th]].

\bibitem{Delduc:2014kha} 
F.~Delduc, M.~Magro and B.~Vicedo,
``Derivation of the action and symmetries of the $q$-deformed $AdS_{5} \times S^{5}$ superstring,''
JHEP {\bf 1410}, 132 (2014)
[arXiv:1406.6286 [hep-th]].

\bibitem{Hoare:2014pna} 
B.~Hoare, R.~Roiban and A.~A.~Tseytlin,
``On deformations of $AdS_n$ x $S^n$ supercosets,''
JHEP {\bf 1406}, 002 (2014)
[arXiv:1403.5517 [hep-th]].

\bibitem{Arutyunov:2013ega} 
G.~Arutyunov, R.~Borsato and S.~Frolov,
``S-matrix for strings on $\eta$-deformed AdS5 x S5,''
JHEP {\bf 1404}, 002 (2014)
[arXiv:1312.3542 [hep-th]].

\bibitem{Araujo:2017jkb} 
T.~Araujo, I.~Bakhmatov, E.~Ó.~Colgáin, J.~Sakamoto, M.~M.~Sheikh-Jabbari and K.~Yoshida,
``Yang-Baxter $\sigma$-models, conformal twists, and noncommutative Yang-Mills theory,''
Phys.\ Rev.\ D {\bf 95}, no. 10, 105006 (2017)
[arXiv:1702.02861 [hep-th]].

\bibitem{Araujo:2017jap} 
T.~Araujo, I.~Bakhmatov, E.~Ó.~Colgáin, J.~i.~Sakamoto, M.~M.~Sheikh-Jabbari and K.~Yoshida,
``Conformal twists, Yang–Baxter $σ$-models and holographic noncommutativity,''
J.\ Phys.\ A {\bf 51}, no. 23, 235401 (2018)
[arXiv:1705.02063 [hep-th]].

\bibitem{Araujo:2017enj} 
T.~Araujo, E.~Ó Colgáin, J.~Sakamoto, M.~M.~Sheikh-Jabbari and K.~Yoshida,
``$I$ in generalized supergravity,''
Eur.\ Phys.\ J.\ C {\bf 77}, no. 11, 739 (2017)
[arXiv:1708.03163 [hep-th]].


\bibitem{Kameyama:2014vma} 
T.~Kameyama and K.~Yoshida,
``A new coordinate system for $q$-deformed AdS$_{5} \times$ S$^5$ and classical string solutions,''
J.\ Phys.\ A {\bf 48}, no. 7, 075401 (2015)
[arXiv:1408.2189 [hep-th]].

\bibitem{Roychowdhury:2016bsv} 
D.~Roychowdhury,
``Multispin magnons on deformed $ AdS_{3}\times S^{3} $,''
Phys.\ Rev.\ D {\bf 95}, no. 8, 086009 (2017)
[arXiv:1612.06217 [hep-th]].


\bibitem{Hernandez:2017raj} 
R.~Hernandez and J.~M.~Nieto,
``Spinning strings in the $\eta$-deformed Neumann-Rosochatius system,''
Phys.\ Rev.\ D {\bf 96}, no. 8, 086010 (2017)
[arXiv:1707.08032 [hep-th]].


\bibitem{Barik:2018haz} 
S.~P.~Barik, K.~L.~Panigrahi and M.~Samal,
``Spinning pulsating strings in $(AdS_5 \times S^5)_{\varkappa }$,''
Eur.\ Phys.\ J.\ C {\bf 78}, no. 4, 280 (2018)
[arXiv:1801.04248 [hep-th]].



\bibitem{Arutynov:2014ota} 
G.~Arutyunov, M.~de Leeuw and S.~J.~van Tongeren,
``The exact spectrum and mirror duality of the $(\text{AdS}_5{\times}S^5)_\eta$ superstring,''
Theor.\ Math.\ Phys.\  {\bf 182}, no. 1, 23 (2015)
[Teor.\ Mat.\ Fiz.\  {\bf 182}, no. 1, 28 (2014)]
[arXiv:1403.6104 [hep-th]].

\bibitem{Khouchen:2014kaa} 
M.~Khouchen and J.~Kluson,
``Giant Magnon on Deformed AdS(3)xS(3),''
Phys.\ Rev.\ D {\bf 90}, no. 6, 066001 (2014)
[arXiv:1405.5017 [hep-th]].


\bibitem{Banerjee:2014bca} 
A.~Banerjee and K.~L.~Panigrahi,
``On the rotating and oscillating strings in (AdS$_{3}$  x S$^{3}$)$_{\kappa}$,''
JHEP {\bf 1409}, 048 (2014)
[arXiv:1406.3642 [hep-th]].





    \end{thebibliography}
\end{document}